# Zero Botnets: An Observe-Pursue-Counter Approach

Jeremy Kepner[1], Jonathan Bernays[1], Stephen Buckley[1], Kenjiro Cho[2], Cary Conrad[3], Leslie Daigle[4], Keeley Erhardt[1], Vijay Gadepally[1], Barry Greene[5], Michael Jones[1], Robert Knake[6], Bruce Maggs[7], Peter Michaleas[1], Chad Meiners[1], Andrew Morris[8], Alex Pentland[1], Sandeep Pisharody[1], Sarah Powazek[9], Andrew Prout[1], Philip Reiner[9], Koichi Suzuki[10], Kenji Takahashi[11], Tony Tauber[12], Leah Walker[9], Douglas Stetson[1]

[1]Massachusetts Institute of Technology, [2]Internet Initiative Japan, [3]SilverSky, [4]Global Cyber Alliance, [5]Akamai, [6]Harvard University, [7]Duke University, [8]GreyNoise, [9]The Institute for Security and Technology, [10]EDB, [11]NTT, [12]Comcast

## Abstract

Adversarial Internet robots (botnets) represent a growing threat to the safe use and stability of the Internet. Botnets can play a role in launching adversary reconnaissance (scanning and phishing), influence operations (upvoting), and financing operations (ransomware, market manipulation, denial of service, spamming, and ad click fraud) while obfuscating tailored tactical operations. Reducing the presence of botnets on the Internet, with the aspirational target of zero, is a powerful vision for galvanizing policy action. Setting a global goal, encouraging international cooperation, creating incentives for improving networks, and supporting entities for botnet takedowns are among several policies that could advance this goal. These policies raise significant questions regarding proper authorities/access that cannot be answered in the abstract. Systems analysis has been widely used in other domains to achieve sufficient detail to enable these questions to be dealt with in concrete terms. Defeating botnets using an observe-pursue-counter architecture is analyzed, the technical feasibility is affirmed, and the authorities/access questions are significantly narrowed. Recommended next steps include: supporting the international botnet takedown community, expanding network observatories, enhancing the underlying network science at scale, conducting detailed systems analysis, and developing appropriate policy frameworks.

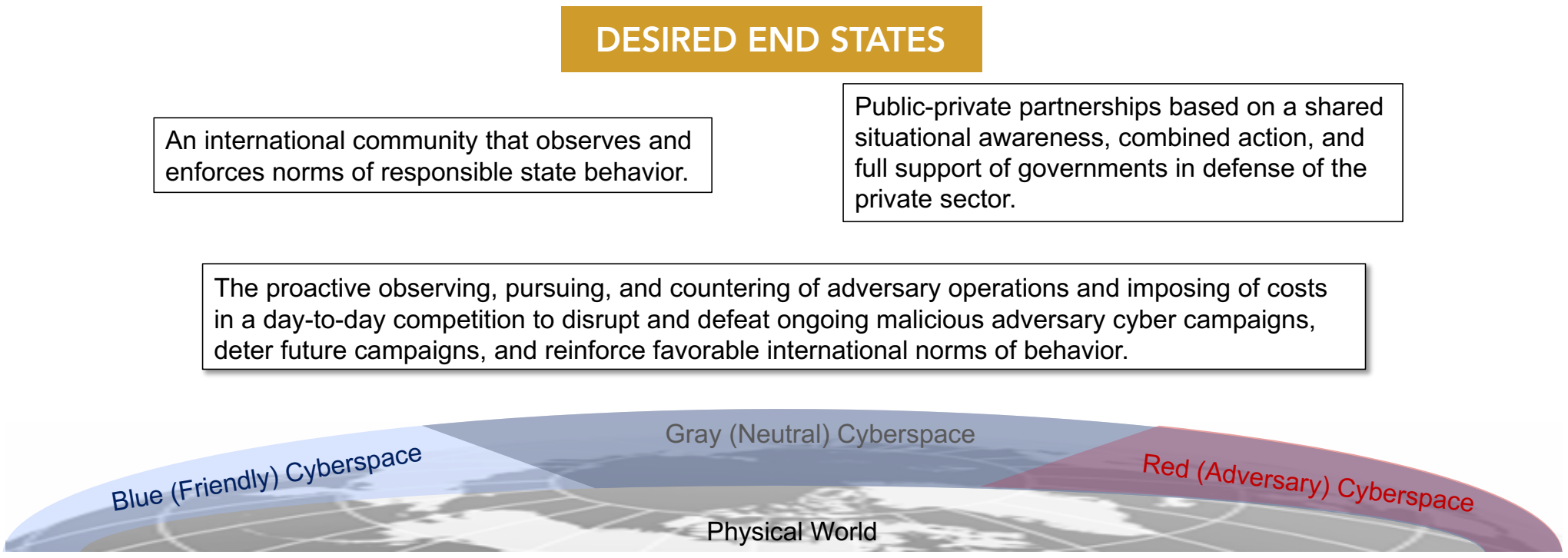

Solarium Report [1]

## Introduction

Botnets are a growing scourge on the Internet. Botnets can play a role in launching adversary reconnaissance (scanning and phishing), influence operations (upvoting), and financing operations (ransomware, market manipulation, denial of service, spamming, and ad click fraud) while obfuscating tailored tactical operations [2][3][72]. The Internet, though revolutionary, remains insecure, and increasing the number of devices on the Internet increases the potential attack surface. This discrepancy has allowed adversarial activity to flourish. Passive devices represent a third of all infected mobile devices in 2020, a 100% increase from 2019 [4], contributing to a situation where adversarial botnets account for a quarter of Internet traffic at some websites (Figure 1)[5].

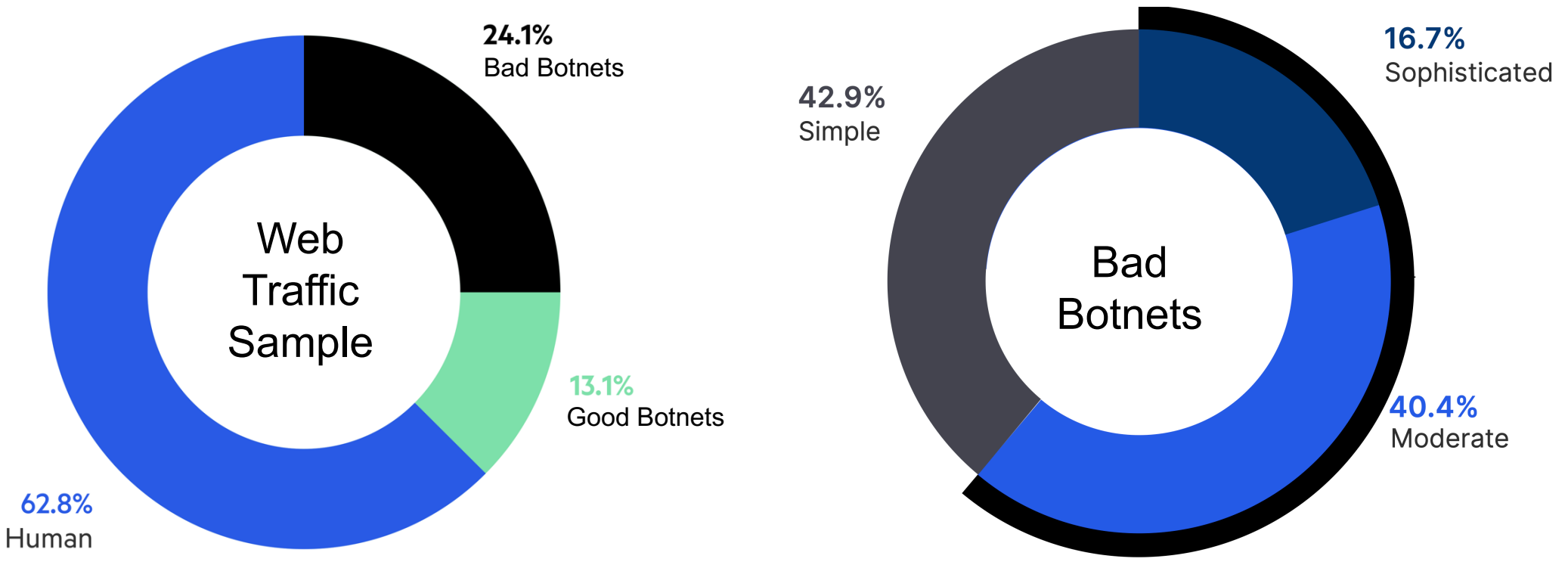

*Figure 1: **Bad Botnets**. Advanced Persistent Botnets (APBs) continue to plague some websites. APBs cycle through random IP addresses, enter through anonymous proxies, change their identities, and mimic human behavior (adapted from [5])*

In an industry where the existence of botnets is often thought of as irreversible, the Council on Foreign Relations (CFR) published "Zero Botnets: Building a Global Effort to Clean Up the Internet" advocating an aspirational zero-tolerance policy for botnets [2]. Describing cyberspace as "the old American Wild West, with no real sheriff and with botnets as the outlaws with guns" the report focused on changing the distribution of cybersecurity responsibility. Most critical to this redistribution of responsibility was the stated need to establish "the principle that states are responsible for the harm that botnets based within their borders cause to others". On the service provider side, the authors urged that "Internet service providers should hold each other accountable for the bad traffic leaving their networks".

Admirable work is ongoing in the non-profit and private sector, with various actors keeping the lights on in the Internet through collecting data containing threat information, taking down botnets, and operating botnet sinkholes [6][7][8][9][10][11]. However, non-governmental actors cannot clear the Internet of botnets on their own. For that, international government coordination is needed. Additionally, there are challenges when relying on private companies to police freedom of speech and freedom of access on the Internet. Simply calling for governments to play a broader role in botnet takedowns is not sufficient. At the moment, there is a lack of prioritization for governments to take down or disrupt botnets that are not engaged in significant fraud or wiretapping.

Governments have treated cyberspace as a warfighting domain for years, but are just starting to address it in the whole-of-domain manner found in land, sea, undersea, air, and space. Governments do not leave sea defense to private fisheries, nor should they leave cyberspace defense to private technology companies. There is an abundance of resources that governments can bring to the fight should they have the right tools and the proper permissions to do so. This paper seeks to present one possible approach to improve the public-private discussion for better network defense. Too often policy discussions are framed by abstract ideas of technical abilities. By providing a specific, feasible, notional architecture, this paper provides the policy debate with the tools to significantly narrow policy questions. Botnets, because of certain distinctive features, provide an opportunity to consider how threats are observed and tracked in the Internet in the context of a specific threat.

The standard approach to protecting existing domains relies on three supporting elements: security, defense, and deterrence (see Figure 2). This paper proposes technical means to address the lack of cyber defense by both private and public actors in regard to botnets. The solution derived, after applying the well tested method of systems analysis (a method used in deriving defense systems for the other domains as well), is a network-based defense, with the precise objective of taking down botnets through an observe-pursue-counter approach. Such a proposal is not novel, having been used in other domains, but applying this approach to botnets requires addressing the particular problem of how to observe cyberspace. Observation in cyberspace is technologically challenging and resource intensive but is critical to understanding where to block and where to shut adversaries down. This approach is one that can be expanded to government and non-government groups, bringing a legitimate technical framework to the policy and technical challenges of ridding the internet of botnets.

## Policy Considerations

The observe-pursue-counter approach calls for the collection and aggregation of sufficient network traffic meta data to identify malicious activity. Such an approach could identify botnets in their early stages of formation when disruption could potentially be easier. It can also be used to inform campaigns to takedown larger, more dangerous botnets. The observe-pursue-counter approach, however, could raise concerns over privacy, warrantless surveillance by government, and other civil liberty concerns. Our analysis is technical and does not address these concerns directly. Our analysis informs the feasibility, cost-benefit trade-offs, and narrows the policy scope of building such an architecture for botnet detection at scale. Our analysis does not assume that governments play a central role in its operation. Business models could be developed that would incentivize an entirely private sector approach. Alternatively, a non-profit organization could be charged with managing the effort. The approach also is does not assume fully monitoring all botnet traffic. A sufficient view could likely be obtained by a series of observatories acting cooperatively and with the consent of their users.

## Background

Botnets come in many forms. A botnet (short for “robot network”) is a network of computers infected by malware that are under the control of an entity, known as the “bot-herder.” A distinguishing feature of botnets is the scale, which may comprise millions of nodes.

Botnets, because of their scale, can amplify other malicious attacks. The centralized control of millions of infected nodes allows botnets to be used in distinctive ways, including what is called a Distributed Denial of Service attack, where all the machines in a botnet flood a host or a region of the network with the goal of disrupting service.

The machines that are part of a botnet are sometimes called “bots”. The term “bot” can be ambiguous and used to describe any program that operates without direct supervision of a human being. “Chatbots” try to have conversations with humans, using AI technology. Some bots of this sort may do manipulative things, such as the “social bots” that join applications such as Facebook and Twitter, with the goal of influencing, upvoting, and the like. Our focus is botnets not bots.

At their essence botnets are distributed computing infrastructure. Adversarial botnets are manifest by their behavior that may include sending undesired communications and exploiting unconsenting computing systems. The architecture of adversarial botnets components often includes:

**Botherder** (botnet shepherder): the entity controlling the botnet.

**BotCC** (botnet command & control): systems that receive direction from the botherder and coordinate the larger botnet.

**Botnet Clients**: usually unconsenting computing systems that have been compromised with botnet malware so as to receive instructions from the botCC to achieve the objective of the botherder and/or spread to other systems.

**Victims**: computing systems receiving undesired communications from the botnet clients.

Botnets create distinctive traffic patterns that can be detected in the Internet. Botnets require complex command and control mechanisms. As the machines that make up the botnet are subverted and turned into clients, those machines must report in and then must stand by for instructions. Many attacks (independent of what the form of the attack) have distinctive traffic patterns.

Large botnets are capable of significant damage. Botnet client level mitigations are manyfold and include continuous software updates, use of two-factor authentication, and changing factory-set device passwords. Taking down the botnet is often the ultimate objective of cyber defenders and would play a significant role on the way to achieving the aspirational goal of zero botnets. The “botnet takedown” community consists of many entities working together to disable botnets. There are many botnet takedown scenarios, but this months-to-years long process often includes the following steps [12][13]

**Victim Identification**: analysis of a variety network artifacts reveals the systems the botnet is victimizing.

**Client Identification**: analysis of a variety network artifacts reveals botnet clients and leads to the botnet malware running on the clients.

**Malware Analysis**: exploration of the operation of the malware in combination with network traffic artifacts reveals how the botnet spreads and communicates with the botCC.

**Client Patching**: fixing client software and adding appropriate signatures to network security systems remediates identified botnet clients.

**BotCC Sinkholing**: seizure of botCC domain names prevents the botherder from controlling botnet clients.

Botnet takedowns play a key role in defending the Internet and involve many parallel components to the takedown stage, including the use of the courts. These may be slow, but it seems to have been effective in many cases. Increasing the pace and reducing the time of botnet takedowns is an important step toward achieving the aspirational zero botnet goal. A high-level assessment of botnets and botnet takedowns suggests a number of acceleration opportunities. Most notably, existing network observatories and outposts demonstrate that the communications among botherders, botCCs, botnet clients, and victims are readily observable from the appropriate vantage points. Early detection of botnet communications allows mitigations to be pursued when botnets are smaller and before they have inflicted significant damage. Accordingly, the zero botnets aspirational goal naturally lends itself to an observe-pursue-counter approach that is a hallmark of effective defense used in a wide range of mature domains (land, sea, undersea, air, and space)[14].

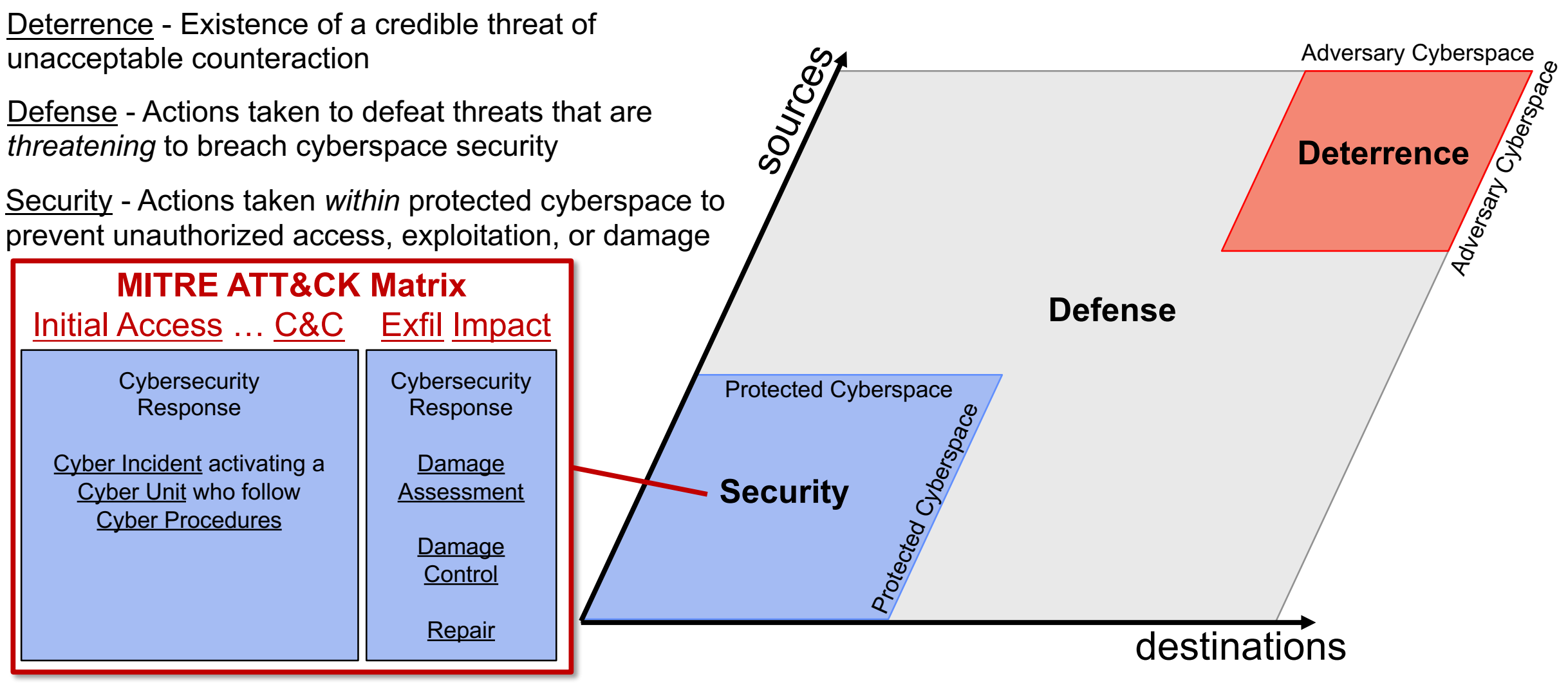

*Figure 2: **Security, Defense, Deterrence**. Source-destination traffic matrix view of cyberspace security, defense, and deterrence using standard domain terminology [66]. Cybersecurity is well characterized by the ATT&CK (Adversarial Tactics, Techniques, & Common Knowledge) paradigm [17].*

A key step in the system analysis process is the appropriate representation of the domain in such a way that is mutually useful to both decision makers and practitioners. The traffic matrix view, where rows represent sources of network traffic and columns represent network traffic destinations is one such generally accepted approach [15]. Furthermore, matrix mathematics (linear algebra) is unaffected by row and column reorderings that come about from anonymization, which means that there are algorithms on traffic matrices that can work on anonymized data [16].

Leveraging the lessons learned from other domains requires contextualizing cyber in broad terms of defense systems analysis. Specifically, the generally accepted definitions of security, defense, and deterrence (see Figure 2). Security covers actions taken within protected cyberspace to prevent unauthorized access, exploitation, or damage. Defense refers to actions taken to defeat threats that are threatening to breach cyberspace security. Deterrence is the existence of a credible threat of unacceptable counteraction. Within the cyber domain, security is the most mature, and is effectively described by the MITRE ATT&CK (adversarial tactics, techniques, & common knowledge) matrix and corresponding actions therein [17]. Deterrence has also received significant resources and is rapidly evolving. Cyber defense of the type that has been most effective in other domains has received disproportionately underinvestment so that modest investments in cyber defense are likely to yield disproportionately positive returns.

## Related Work

Network based defense has long been recognized as offering many benefits [18][19]. Network-based systems observe the traffic generated between multiple hosts. They are placed strategically at ingress/egress points to capture the most relevant or risky traffic. Suspicious behaviors that may be flagged include failed connection attempts, failed domain name server (DNS) requests, web connections to blacklisted sites, and the use of randomized domain names. Network-wide monitoring provides an overview of all activity in the observable space to aid in detecting behavioral patterns, fluctuations, and group actions. A primary disadvantage of network observations is the sheer volume of data generated daily, requiring greater resources to collect and process. Historically, the volumes of data required have been perceived as so insurmountable that network-based defense has been discounted, particularly in the context of providing the view of the Internet necessary for an observe-pursue-counter approach to botnet takedowns. Fortunately, the advent of more performant AI (artificial intelligence) algorithms, software, hardware, and cloud computing capable of operating on anonymized data has made network-based defense systems with sufficient capability routine in the private sector [20]. While the methods used in the private sector are generally proprietary [21], more recently, open approaches demonstrating many of the required capabilities on anonymized data have been published by the academic research community. Some of these innovations are described as follows.

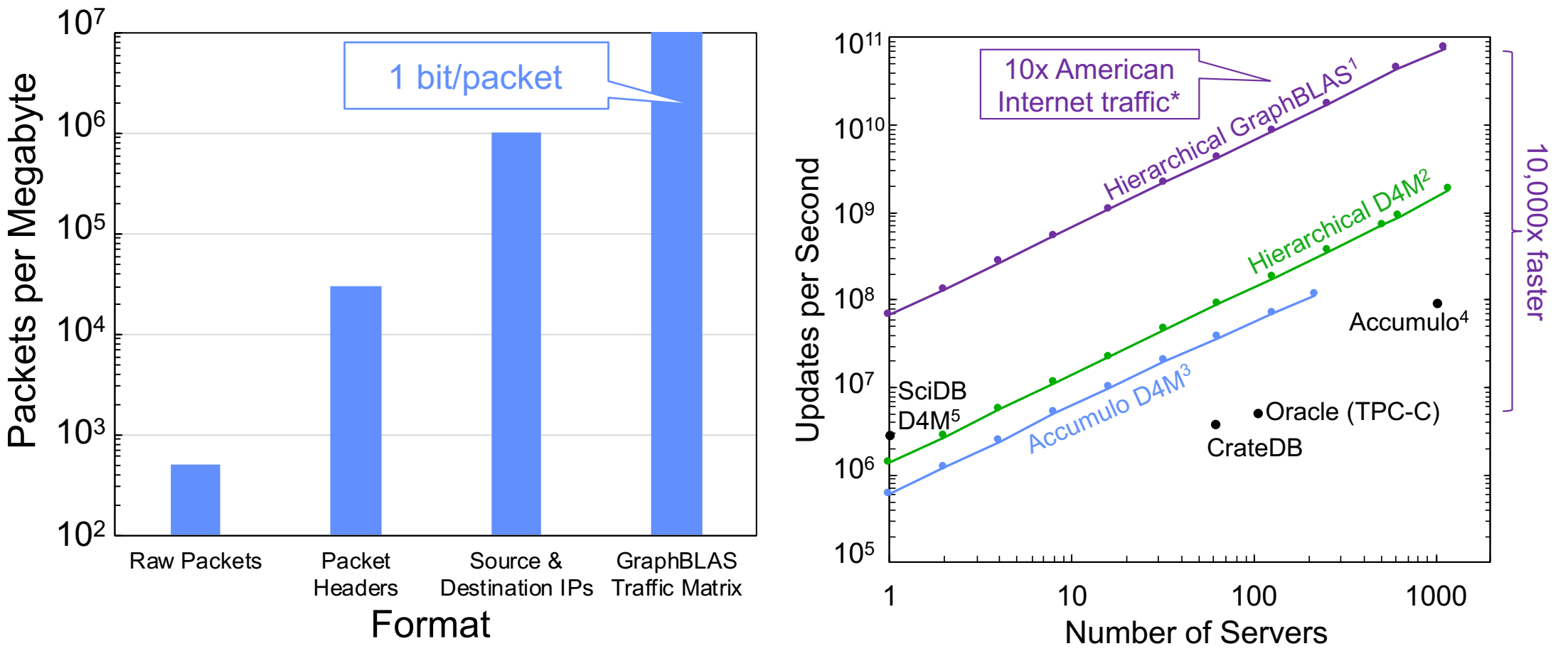

*Figure 3: **Feeds and Speeds for Anonymized Traffic Matrices**. The GraphBLAS.org open standard provides a sparse traffic matrix library demonstrating compression (left) and performance (right) on anonymized traffic matrices consistent with existing proprietary capabilities needed to observe larger-scale networks. *Non-video traffic [22][23], [1][24], [2][25], [3][26], [4][27], [5][28].*

The observe-pursue-counter approach is the foundation of many domain defense systems. The observe component is often the most technologically challenging and resource intensive. Thus, systems analysis usually begins with an assessment of the technological fundamentals necessary for effective observation of the domain. These "feeds and speeds" are typically the volumes of data to be stored and the rates at which they can be processed. The published open state-of-the-art for anonymized network traffic matrices using the GraphBLAS.org standard is shown in Figure 3 and affirms the basic feasibility and claims made by the private sector. A thousand server system with a commodity interconnect can process 10x the non-video traffic of the North American Internet. While such a system may seem large, it is 1% of a typical hyperscale datacenter, of which there are hundreds worldwide.

A common aspect of many recent AI innovations is their reliance on signatures of the phenomena they are tasked with identifying (referred to as supervised machine learning). AI algorithms often require copious amounts of clean training data with clearly marked examples. Cyber security has generally adopted a signature-based approach to detection that has become overwhelmed by the exponential diversity that is readily achievable in modern malware. Even once the signature of new malware is identified, an update must be deployed to pre-existing AI systems to include the new signature - a process which can be slow on many systems and fail to keep pace with the malware's evolution. Other domains face similar challenges. An air defense system that relied on detailed signatures of every aircraft in the world would be impractical. As a result, defense systems in other domains instead use AI to model the background for which there is copious amounts of training data. Using highly accurate background models, anomalies can be readily detected and enriched with additional sensor modalities to allow precise classification. Similar approaches are used for cyber defense in the private sector. For reviews of the significant broader literature in this space see [29][30][31][32][33]. Selected examples from the open literature drawn from the authors' work on anonymized traffic data are presented in Figure 3-Figure 7.

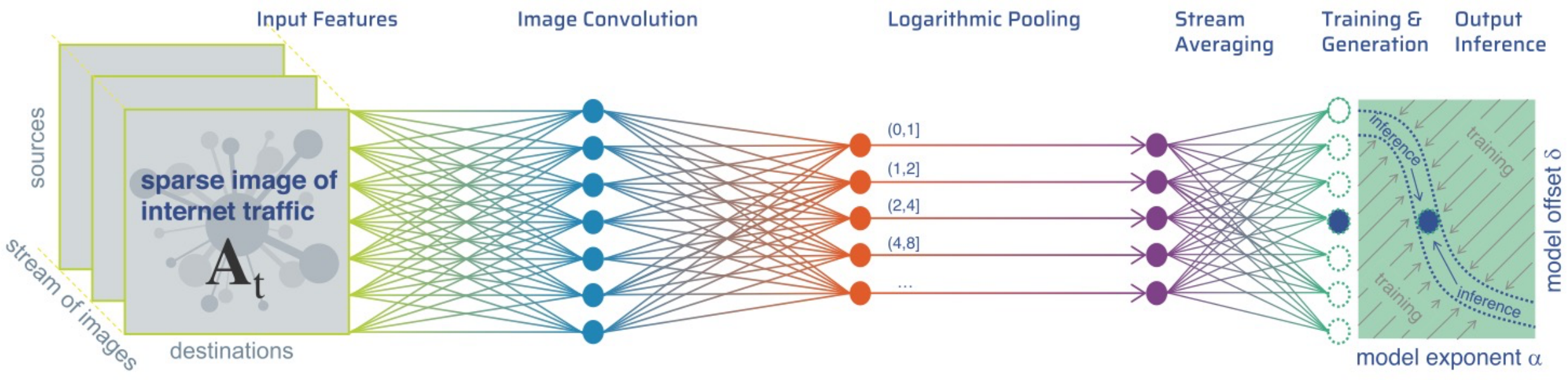

*Figure 4: **AI Background Modelling on Anonymized Traffic**. Treating anonymized network traffic matrices as a stream of sparse images allows standard AI methods to extract accurate features for inferring precise background model parameters $(\alpha, \delta)$ [25]. These models can then be used for anomaly detection.*

For many decades signal processing has been the basis of the detection theory that underpins the observe component of most effective domain defense systems [34] [35][36]. These signature-less approaches compute accurate models of the background signals in the data. Comparing observations with these background models is an effective way to detect subtle anomalies. Figure 4 is an example of the modern AI equivalent of this approach applied to anonymized network traffic matrices [25]. Treating the network traffic matrices as a sequence of sparse image enables standard convolutional neural network methods to be used to train accurate models of background network traffic [37][38].

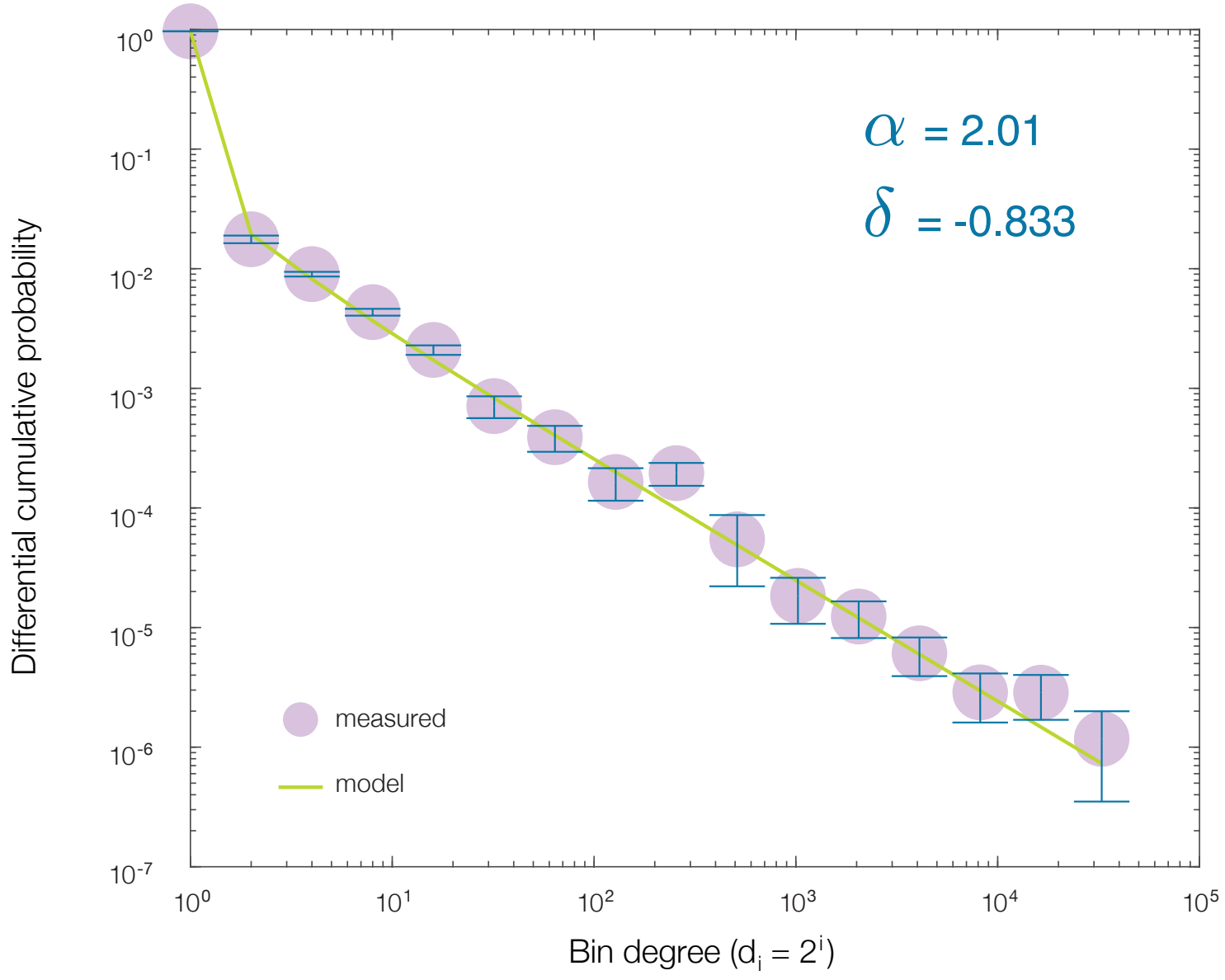

*Figure 5: **Example Observed Background from Anonymized Traffic**. Many network quantities: packets, sources, destinations, links, … exhibit a power-law behavior. With sufficient data and AI processing, accurate network traffic measurements can be made that allow precise background model fitting, in this case $p(d) \propto 1/(d+\delta)^{\alpha}$, that can be used for signature-less anomaly detection on anonymized traffic data [25]. Similar background models are widely used in proprietary network defense solutions.*

The Gaussian or normal distribution specified by a mean and variance is a standard background model used in many domains. One of the most significant early discoveries in the field of Network Science is that the probability distribution of many network quantities: packets, sources, destinations, links, … exhibit a power-law [39][40]. With sufficient traffic, these distributions can be measured accurately enough to train high precision models of the background (see Figure 5). These power-law distributions and their parameters can be computed entirely from anonymized traffic matrices [25]. Similar background models are widely used in proprietary network defense solutions.

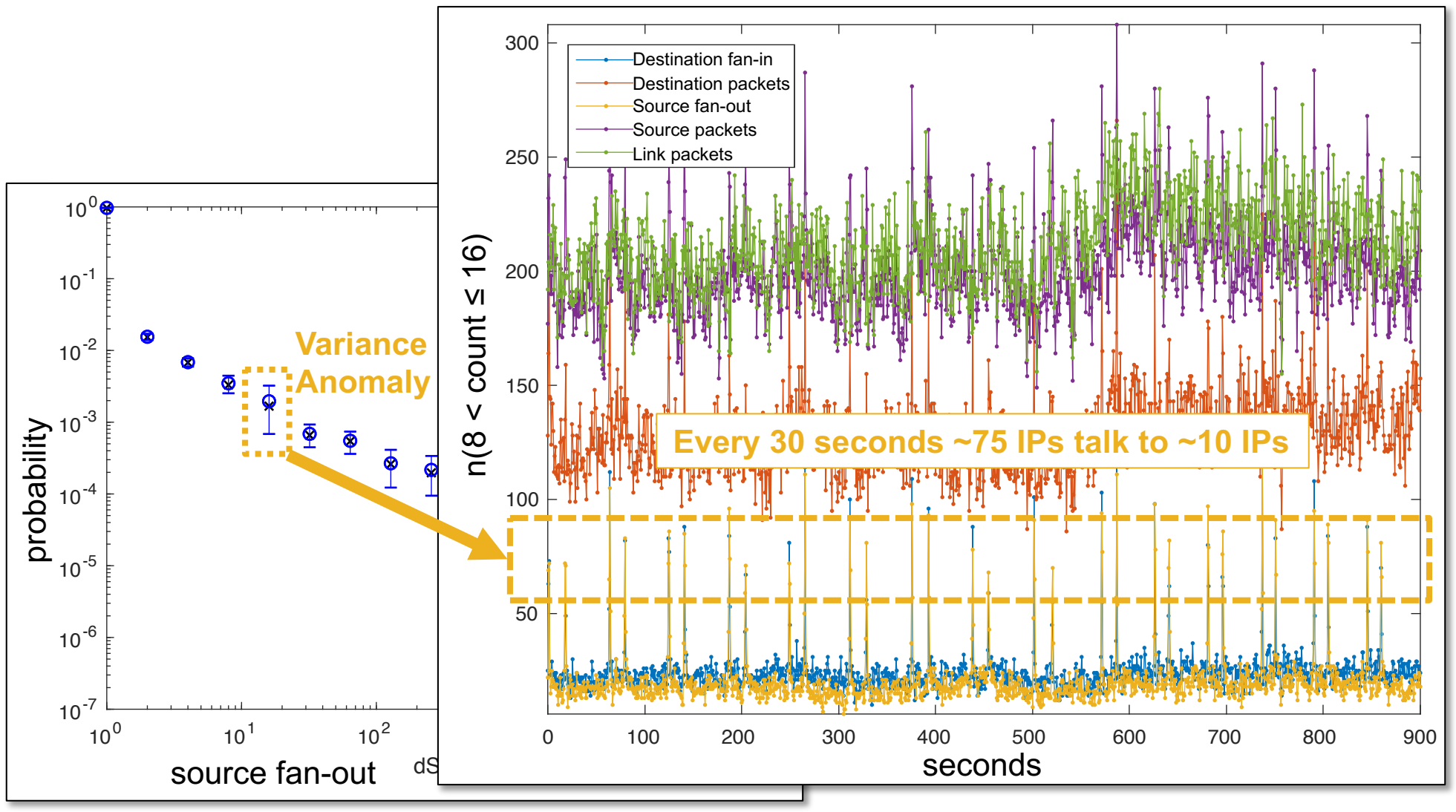

*Figure 6: **Anomaly Detection on Anonymized Traffic**. Accurate measured backgrounds allow for the detection of small anomalies on anonymized traffic data. In this case an anomaly in the variance of sources with 9 to 16 destinations reveals likely botCC traffic [41]. Similar anomaly detection methods are widely used in proprietary network defense solutions.*

Figure 6 provides an example of how an accurate background measurement can be used to detect anomalies. In this case, the probability distribution on the number of unique destinations each source is connecting to (the source fan-out) shows a higher variance for sources with 9 to 16 destinations. Zooming in on the corresponding time-series data reveals a regular spike in activity where a different set of 75 sources are talking to 10 sources [41]. This pattern is characteristic of botCC traffic and could be used to map out a botnet. Figure 6 is one of the many examples of the type subtle anomalies that are routinely detected in proprietary network defense solutions using accurate background models. In this example, the anomalies and their parameters can be computed entirely from anonymized traffic matrices.

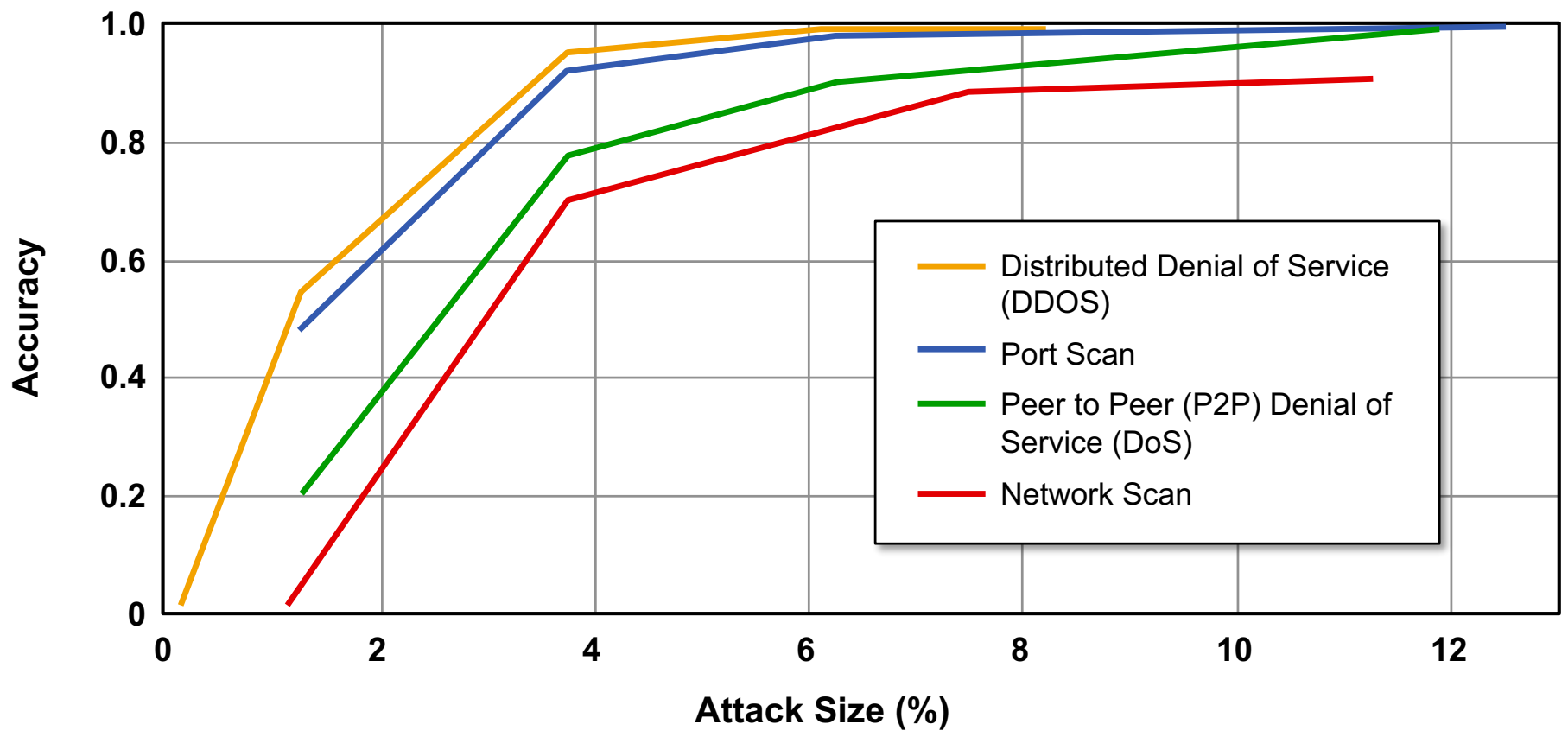

*Figure 7: **Anomaly Classification on Anonymized Traffic**. Classification results of an AI based anomaly classifier on anonymized traffic data as function of the attack size as a percentage of the total traffic [42]. Similar AI anomaly classification methods are widely used in proprietary network defense solutions.*

Background models of the dynamics of the data-defined network (analogous to the hardware and software defined network) are particularly useful because when done correctly they can be much more sensitive than simple traffic models or nodal characteristics [43]. Functional networks (like botnets) necessarily show coordination in time, in addition to displaying atypical patterns in connectivity. A good example of this is that when establishing a botCC, botnet malware often "beacons" by sending regular messages to potential command and control servers. Detection of these patterns of time coordination ("coincidences") in metadata have much better statistical leverage than detection of atypical traffic patterns. The computational cost of characterizing network dynamics appears, at first glance, to be prohibitive. However, good approximations can be obtained in a time that is roughly N log(N), where N is the number of nodes [44]. This method has been used in a wide variety of networks [45], and is used operationally by the Israeli and Singapore defense forces to analyze large numbers of streams of encrypted data [46]. Most recently it has also been demonstrated that this type of analysis can detect controller or influencer subnetworks, and characterize their effects [47].

Using AI to create a background model enables signature-less anomaly detection. After an anomaly has been detected it can be passed through a signature-based AI classifier to see if the anomaly matches a previously observed phenomena or is novel [48][49][50][51]. Figure 7 is an open example of this approach that is widely used in proprietary network defense systems [42]. Figure 7 shows how AI can identify the network traffic patterns of several of the most common types of botnet activities: distributed denial-of-service (DDoS), point-to-point denial-of-service (P2P DoS), port scanning, and network scanning. In this example, the classifications can be computed entirely from anonymized traffic matrices. In this type of system deanonymization might only be required when the pursue-counter stages are invoked. Only deanonymizing sources that are likely inside a botnet can lead to a billion-fold reduction in the amount of deanonymized data required. For example, compare total packets in Table 1 ($10^{18}$/year) with the size of botnets in Table 2 ($10^{7}$/year).

| Detection Requirements |
| --- |
| The observe component is often the most technologically challenging and resource intensive component of the observe-pursue-counter approach. Researchers continually strive to achieve detection outcomes with higher accuracies and lower false-alarms rate. Detection research is essential for progress and should be strongly supported. However, "perfection" can be the enemy of "good enough". Systems analysis provides a mechanism for answering these difficult detection questions. Specifically, once the foundational science and engineering of the observe component is understood, the detailed detection requirements are driven by the pursue-counter elements of the system. If a particular counter is relatively benign then the false-alarm rate can be higher. In contrast, if a counter could be potentially very disruptive, then the false-alarm rate must be lower. Using this approach, it is possible to conduct an effective systems analysis and develop a reasonable systems architecture without precise specification of every detection detail. |

## Methodology and Architecture

Defense is significantly aided if the time and place of attack are known. To achieve this foresight, domain defense relies on an observe-pursue-counter architecture (aka detect-handoff-intercept). Networks are a distinguishing characteristic of the cyberspace domain. Malware without a network is a rare threat. Networks without malware are still a major threat. It is natural that this approach would focus on networks.

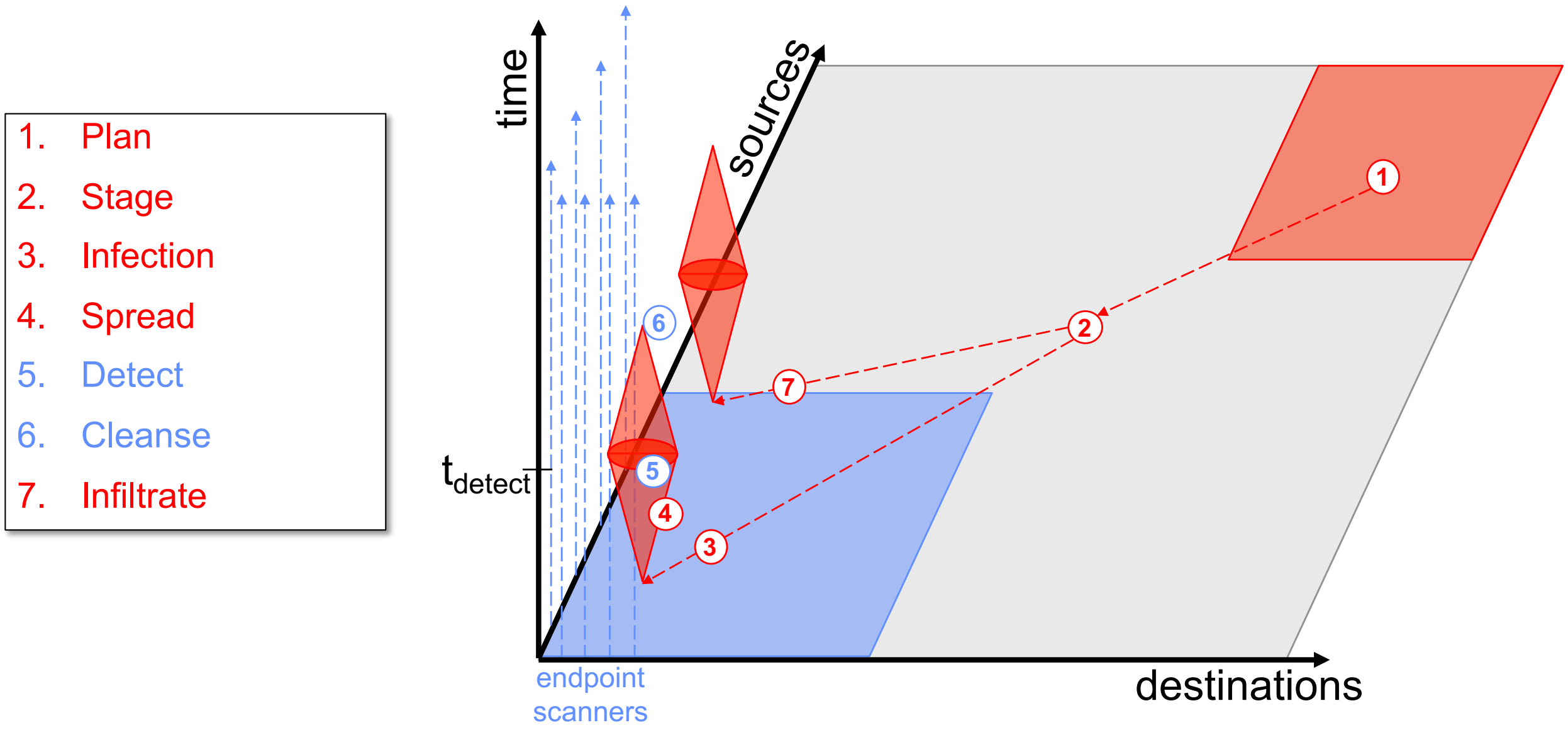

*Figure 8: **Notional Botnet Growth (Current)**. A typical growing botnet is detected (5) and cleansed (6) long after initial infection (3) when the spread (4) is eventually uncovered by an endpoint scanner equipped with the appropriate signature. This lag allows additional spread to occur (7).*

Figure 8 illustrates the typical growth and spread of a botnets and other malware and how the spread is mitigated using current cybersecurity means. Because growth is via a network, a standard security approach is most likely to engage after significant spread has occurred. The spread starts with planning in the adversarial domain (1), followed by staging in neutral (gray) space (2), from which infection into the protected space can be achieved via a variety of techniques (3). Once inside a protected domain, spreading begins and expands the footprint of adversarial capability (4). Once endpoint scanners have been loaded with the appropriate signature the botnet is detected at time $t_{detect}$ (5) which is often months after the initial staging of the adversarial activity. Upon detection, widespread cleansing can take place (6). Meanwhile the process continues to repeat itself indefinitely elsewhere throughout the network (7).

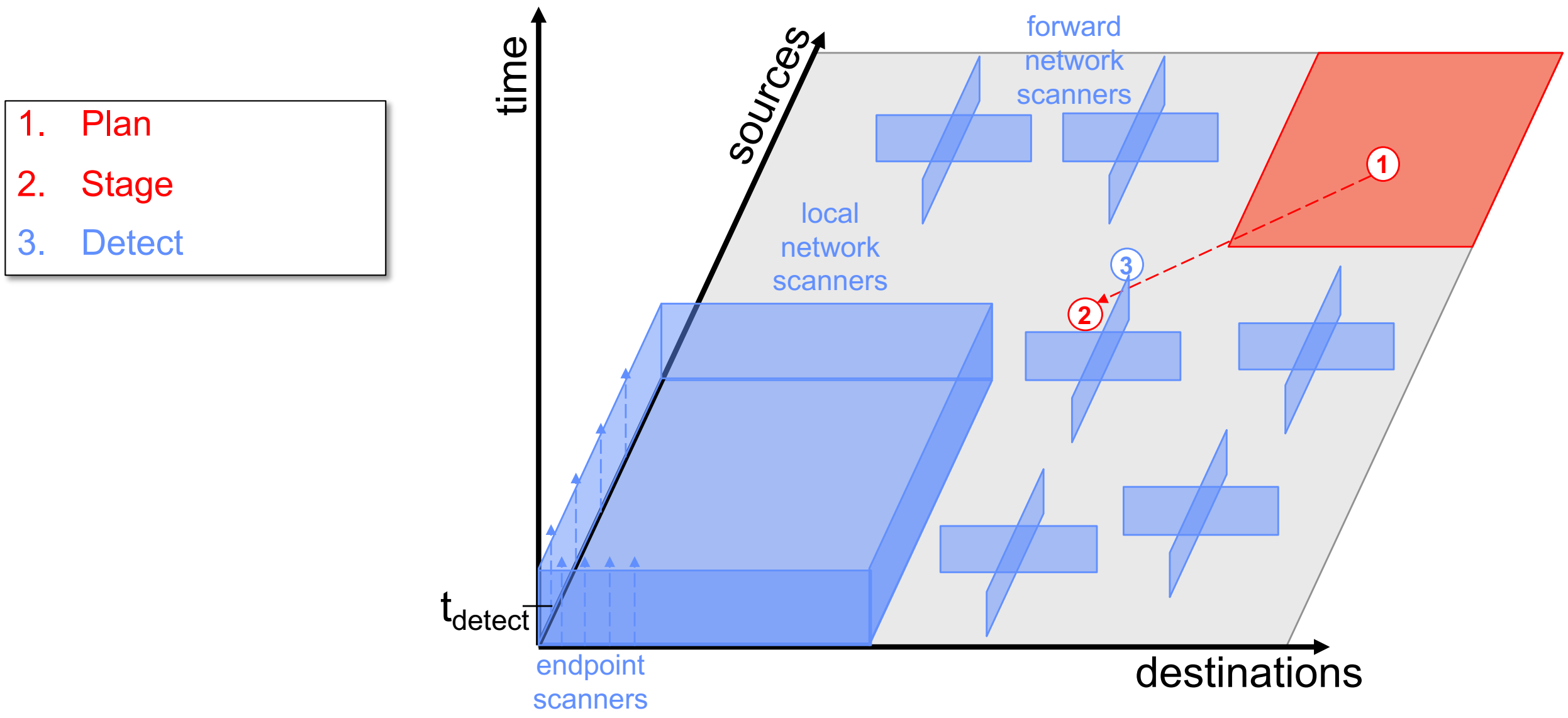

*Figure 9: **Notional Botnet Growth (Future)**. A collaborative observe-pursue-counter approach with sufficient observation of network traffic can detect (3) adversary activity during the planning stage (2) when it is much smaller and easier to mitigate.*

The key parameter for determining the effectiveness of most domain defense system is $t_{detect}$. The smaller $t_{detect}$ the more effective the system will be. The current large scale and impact of botnets is consistent with current high value of $t_{detect}$. Likewise, if $t_{detect}$ could be significantly reduced a corresponding reduction in the scale and impact of botnets would be expected. A standard way to shorten $t_{detect}$ is to "defend forward" so that earlier stages in the process are visible. Figure 9 illustrates a collaborative architecture with sufficient view to observe the earliest stages of a botnet. Specifically, detection (3) of the staging step of a botnet (2) outside of the protected space. While such an architecture is easy enough to describe theory, systems analysis is required to affirm the practical feasibility of such an approach.

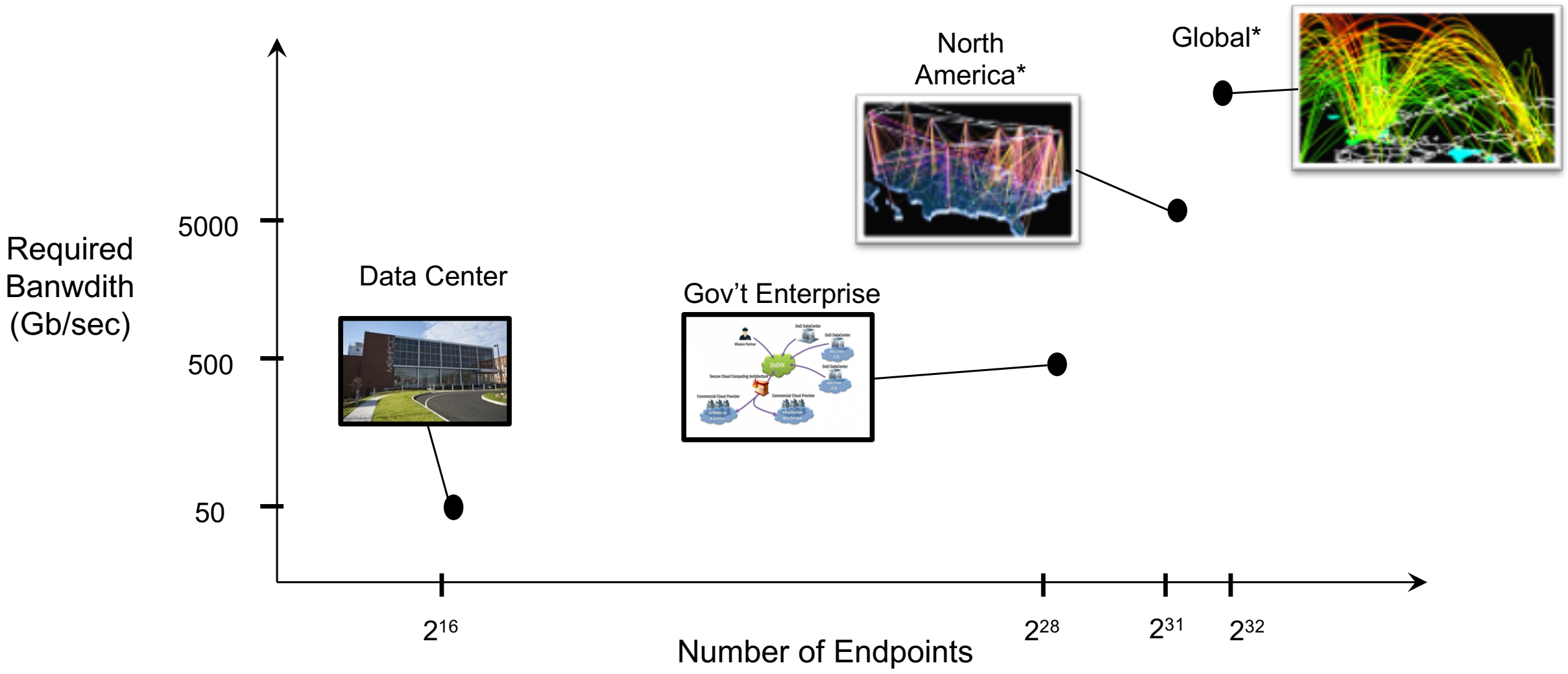

*Figure 10: **Relevant Scales**. Cyber defense can be implemented at many different scales. Each scale has different traffic loads and resource constraints [52][53]. *Non-video traffic.*

An important step in defense systems analysis is determining the relevant scales of the domain to be defended. An architecture that can be implemented at multiple scales provides the necessary opportunities to learn while doing. Architectures that are only effective if the entire domain is covered, are higher risk and require more up-front resources. Figure 10 shows the estimated number of endpoints and bandwidth needed to observe a significant fraction of the traffic for a typical data center, government enterprise network, all of North America non-video traffic, and worldwide non-video traffic. For this analysis, video is excluded because video does not play a significant role in botnet activity, and, by design, is relatively easy to identify and distinguish in network traffic.

*Table 1: **Estimated Traffic at Different Scales**. Baseline bandwidth, storage capacity, and costs for a view of traffic at different scales suggesting the feasibility of storing and processing data at these scales. [1][22][23]*

| | Data Center | Gov't Enterprise | N. American Internet [80% video][1] | Global Internet [89% video, 70% CDN][1] |
|---|---|---|---|---|
| Link | 20 Gb | 500 Gb | - | - |
| Duty factor | 50% | 50% | - | - |
| Number of devices | 1M | 10M | 10G[1] | 20G[1] |
| Data bandwidth | 1.25 GB/s (40 PB/yr) | 30 GB/s (1 EB/yr) | 30 TB/s (1 ZB/yr)[1] | 100 TB/s (5 ZB/yr)[1] |
| Packet rate (1 KB/packet) | 1.25 M/s (40 T/yr) | 40 M/s (1 P/yr) | 7 G/s (250P/yr) | 20 G/s (1E/yr) |
| Anonymized GraphBLAS matrix data rate | 150 KB/s (5 TB/yr) | 20 MB/s (125 TB/yr) | 1 GB/s (30 PB/yr) | 3 GB/s (150 PB/yr) |
| Min data storage cost ($100/TB) | $500/yr | $12,500/yr | $3M/yr | $15M/yr |

Table 1 estimates the required bandwidth and storage needed to achieve the necessary view of network traffic to provide basic observations at various relevant scales. Fortunately, the recent technical innovations demonstrated in the research community and available in industry indicate that the data rates, data volumes, and corresponding storage costs are accessible at these scales. Table 1 provides an initial affirmation that sufficient observations are technically feasible.

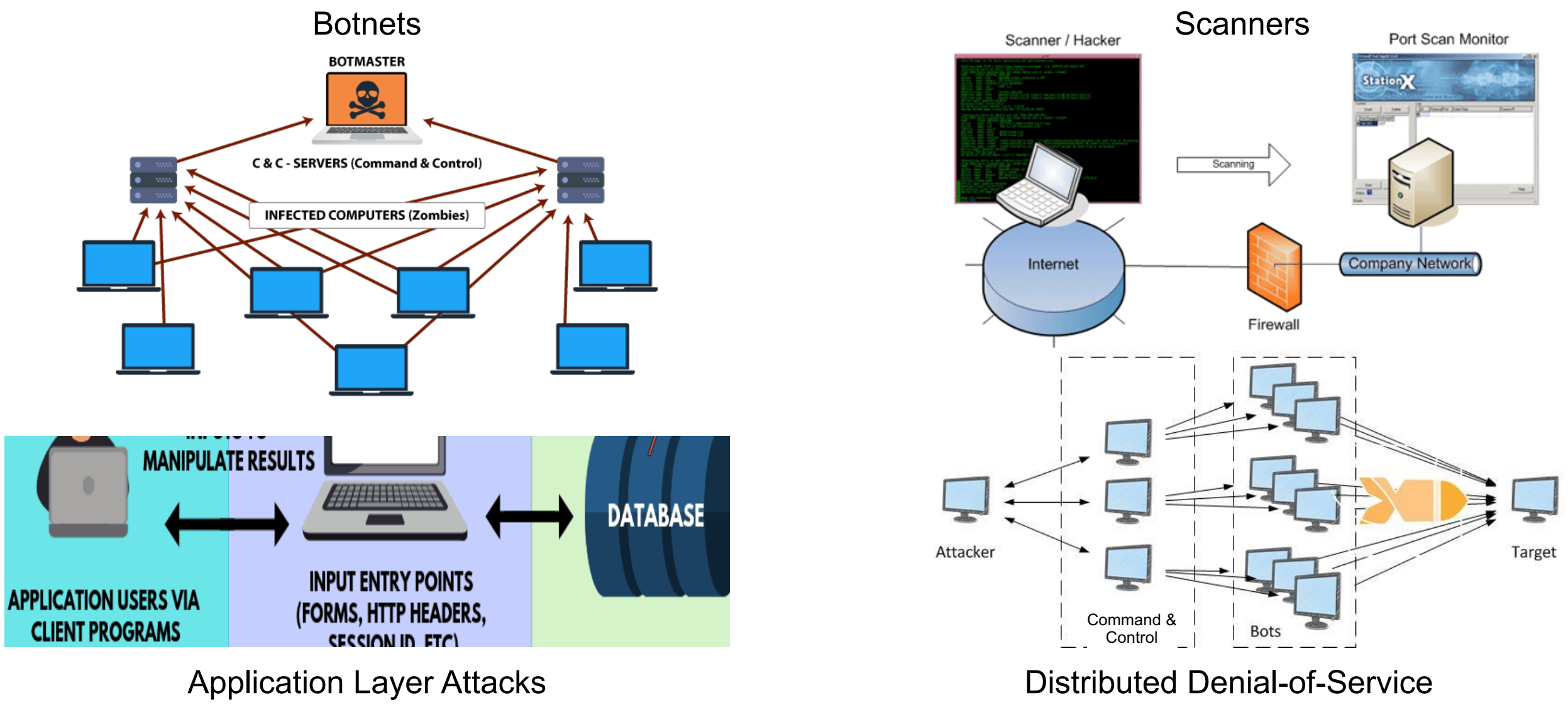

*Figure 11: **Major Adversary Network Activities**. Botnets and their scanners comprise a significant portion of the infrastructure supporting application layer, distributed denial-of-service (DDoS), and other cyberattacks [54][55][56][57]. Eliminating botnets has a manyfold benefit.*

Table 1 establishes the typical scales of overall network activity. It is also important to estimate the scale of the specific phenomena to be detected in the overall traffic. Figure 11 provides examples of major adversary activities on a network. Botnets and their scanners comprise a significant portion of the infrastructure supporting application layer, distributed denial-of-service (DDoS), and other cyberattacks. Table 2 provides estimates of the scale of these activities demonstrating that they are significant and readily observable, while being of sizes and rates that are readily processed.

*Table 2: **Adversary Traffic**. Major adversary network activities at different scales suggesting the detectability and feasibility of these activities. *Non-video traffic, [1][22], [2][23], [3][5], [4][58], [5][59], [6][2], [7][60], [8][61], [9][62]*

| | Data Center | Gov't Enterprise | N. American Internet* | Global Internet* |
|---|---|---|---|---|
| Adversary botnet packets (25%)[3] | 300 K/s (2.5T/yr) | 10 M/s (125T/yr) | 2 G/s (65P/yr) | 6 G/s (250P/yr) |
| New Botnet C&Cs identified | | | 350/month[5] | 1000/month[5] |
| Number of bots | | | 200K[6] | 10M[6] |
| Malicious login attempts | | | 17 G/yr[7] | 42 G/yr[7] |
| SQLi (77% of Web App attacks) | | | 1 G/yr[7] | 4 G/yr[7] |
| Scanner sources[8] | 1.1 M/month | 1.1 M/month | 1.1 M/month | 1.1 M/month[8] |
| Scan packets/IP dest[8] | 117 K/year | 117 K/year | 117 K/year | 117 K/year[8] |
| Scan packets | 3.7 K/s (100 G/yr) | 37 K/s (1 T/yr) | 250 M/s (830 T/yr) | 750 M/s (2.5 P/yr) |
| DDoS Attacks | | | 4.7 M/yr | 9.4 M/yr[1,2,9] |
| DDoS Attack Campaigns | 0.3/yr | 15/yr | 10 K/yr[4] | 20 K/yr[4] |
| DDoS campaign duration[4] | 7.8K sec | 7.8K sec | 7.8K sec | 7.8K sec |
| DDoS bandwidth/campaign[4] | 175 MB/s | 175 MB/s | 175 MB/s | 175 MB/s |
| DDoS packets/campaign (1K packet) | 1.4 T | 1.4 T | 1.4 T | 1.4 T |
| DDoS packets | 10 K/s (400G/yr) | 700 K/s (20 T/yr) | 500 M/s (14 P/yr) | 900 M/s (30 P/yr) |

## Observatories and Outposts

Observations are the centerpiece of an observe-pursue-counter approach in any domain. Network observatories (that share anonymized traffic with researchers) and outposts (that share traffic analysis with customers) are critical. These lookouts provide empirical answers to critical questions, such as

- How big is the Internet?
- How many devices are connected?
- How much traffic is flowing?
- What is the nature of this traffic?

Without answers to these basic questions, policy discussions have a limited quantitative basis for evaluating the scale and impact of proposals.

Fortunately, there are a number observatories and outposts in operation today. A selection of these capabilities is shown in Figure 12. These sites are a mixture of academic, non-profit, and commercial efforts and provide different viewpoints into the network landscape. Some sites lie on gateways of protected space. Others lie along major trunk lines in grey space. And others are honeypots or darks spaces (unassigned locations on the Internet) that see mostly adversarial traffic. Even larger capabilities exist within larger proprietary network entities [63][64][65].

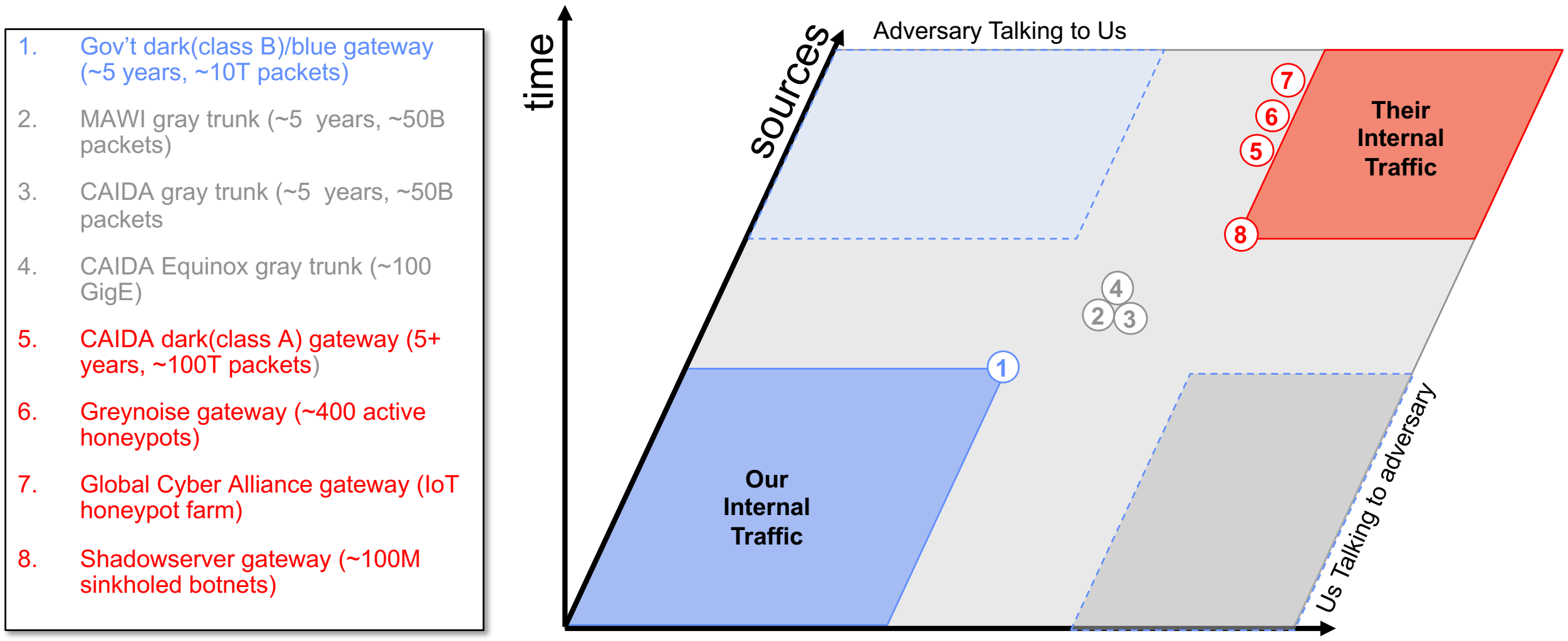

*Figure 12: **Observatories and Outposts**. Examples of current observatories and outposts and their approximate proximity to various network traffic [6][7][8][9][10][11][67]. Expansion of these capabilities would provide a significant improvement in network awareness.*

It is worth highlighting the CAIDA and MAWI observatories that have continued to be pioneers in this field over the past decades and have demonstrated the value and proper handling of these observations. It is generally acknowledged that the broad sharing of network traffic it to be avoided. However, the CAIDA and MAWI model of sharing anonymized traffic header information with registered legitimate entities under appropriate agreements to perform specified research is effective and has become a model for proper data sharing in other domains.

A key design principle in data sharing is co-design [68]. Most data at initial collection are unusable and unsharable. Understanding the data and its purpose are necessary for effective usage and sharing. Identifying and maintaining data curators and key points-of-contacts among entities is critical. Iterative development approaches among selected entities sharing limited subsets of data are an effective way to mature data products, their formats, and anonymization procedures. Avoiding proprietary formats and adopting the simplest widely accepted tabular formats along with using simple file naming and folder structures are significant enablers. Subsequently automating these processes at a data collection site can maximize the benefit to both the site and the broader community.

The aforementioned processes can significantly mitigate data sharing concerns (see Appendix A). Currently, confusion on data sharing liability limits willingness to share data with researchers. Data owners aim for the common denominator of international requirements (US, EU, …). Fortunately, standard practices now exist that meet these requirements, that include

- Data is made available in curated repositories
- Using standard anonymization methods where needed: hashing, sampling, simulation, …
- Registration with a repository and demonstration of legitimate research need
- Recipients legally agree to neither repost a corpus nor deanonymize data
- Recipients can publish analysis and data examples necessary to review research
- Recipients agree to cite the repository and provide publications back to repository
- Repositories can curate enriched products developed by researchers

Funding entities, journals, conferences and professional societies should encourage research conducted under these conditions.

Finally, the burden of making final decisions with regards to information sharing often falls to information security officers (ISOs) within data holding entities. A key to accelerating effective data sharing is enhanced communication between ISOs and subject matter experts (SMEs). ISOs and SMEs have different terminology. ISO sign-off requires confidence in SME data handling practices. An ISO requires basic information to allow data sharing: project, need, location, personnel, duration, ... SMEs often provide highly technical research-oriented descriptions of data that limit ISO surety in SMEs data handling practices and results in ISOs limiting data sharing requests. Better training of SMEs to allow them to more effectively communicate with their ISOs is an important part of an effective data sharing policy (see Appendix B).

## Findings and Discussion

The early stages of development in an operating domain are often characterized by rapid innovation that contrast with established domains. During this period of innovation it is natural to adopt an exceptionalist view that focuses on the how the new domain differs from other domains. As the domain matures and aligns with other domains it becomes easier to consider the new domain within the context of established norms.

A first step for cyber has been alignment with the standard concepts of security (walls-in), defense (walls-out), and deterrence found in other domains (Figure 2). Within these standard concepts are established norms of acceptable security, defense, and deterrence behavior that can be used as models for the cyber domain.

The recognition of the observe-pursue-counter approach to cyber defense further aligns cyber with traditional defense constructs in other domains [1]. [Note: In other domains this construct is often referred to as detect-handoff-intercept.] The use of standard defense constructs further lend themselves to standard defense systems analysis and architecture.

Scale and costs and critical drivers of any system. The observe component is often the most resource intensive. Obtaining the necessary observations of network traffic has been daunting, but advances in compression and processing have brought these within reach (Figure 3). Systems analysis establishes the scale of a domain (Table 1) and the phenomenology of interest (Table 2). This is critical for quantifying the potential impact of threats in an unbiased and rigorous manner. Without such quantification it can be difficult to distinguish significant threats that effect millions from rare threats relying on exceptional trade craft.

Questions about authorities/access are commonplace early in a domains' development. These questions can appear overwhelming as practitioners strive for broader authorities/access to improve observation and provide more accurate detection, but such authorities/access and access are not consistent with standard norms. Systems analysis has been widely used in other domains to achieve sufficient detail to enable these questions to be dealt with in concrete terms. Existing network observatories and outputs have pioneered methods for effective observing and sharing network traffic to enable detection research in a manner consistent with societal norms (Figure 12). Using these data, researchers have also demonstrated that significant progress can be made using AI approaches that operate on anonymized data (Figure 4Figure 7) leading to a potential billion-fold reduction in the amount of deanonymized required. Additionally, recognition that detection requirements are set by the specifics of the pursue-counter components of the defense architecture liberates detection research and narrows policy considerations to specific contexts. Combined, the progress on effective data sharing, anonymized AI algorithms, and pursue-counter driven detection significantly narrows authorities/access questions.

Finally, it has been suggested that the cyberspace lacks the fundamental "physics" found in other domains. The underlaying phenomena in any domain is rarely inherently manifest and must be discovered by painstaking science. Initial observations of the early Internet revealed a variety of new phenomena leading to the establishment of the new field of *Network Science* [69]. Current observations are a million times larger and are calling out for scientific exploration.

## Conclusions and Next Steps

Airlines, air freight companies, and air passengers pay taxes and expect airspace to be defended against adversarial nations. Cruise ships, shipping companies, and sea passengers pay taxes and expect the ocean to be defended against adversarial nations. Telecommunication companies, service providers, and Internet users pay taxes and expect cyberspace to be defended against adversarial nations. Domain defense (land, sea, undersea, air, space, cyber) against adversarial nations is a government responsibility and governments have an important role to play in defending those domains. Exploration and validation of this approach should be a part of the overall cyber strategy, and cannot be left entirely to any single entity, because the needs for scale and data fusion do not sit at the enterprise perimeter, but require making observations on the network as a whole. Along these lines, there are several concrete next steps that can be taken to towards the zero botnets aspirational goal.

Support the International Botnet Takedown Community

There is a small dedicated community fighting this fight today. They should be actively supported [Knake 2021].

Expand Observatories and Outposts

The globe currently depends upon a small dedicated community to operate and maintain current network observatories and outposts. These lookouts are our only means for obtaining consensus empirical answers to critical questions. These capabilities should be significantly expanded.

Enhancing the Underlying Network Science at Scale

Understanding the underlying processes in all fields is discovered by painstaking science. Early efforts on small data sets revealed significant new discoveries and established the field of Network Science. Current observations are a million times larger and are calling out for scientific exploration.

Detailed Systems Analysis and Trade Studies

The preliminary systems analysis of defeating botnets with using an observe-pursue-counter architecture affirmed the basic technical feasibility and has narrowed the authorities/access questions. Detailed systems analysis is required to establish the specifics of potentials architectures and conduct necessary trade-off studies to develop larger efforts. Such analysis can use existing proposed architectures as a starting point (see Figure 13).

Developing an Appropriate Policy Framework for Usage

Which data to collect? By who? For what purpose? Are critical policy questions. Existing efforts establish that significant value can be derived from existing data. Sharing models for anonymized traffic header information with registered legitimate entities under appropriate agreements to perform specified research has become a model for proper data sharing. These efforts provide a significant foundation for developing the appropriate policy framework for a data-driven approach toward the aspirational goal of zero botnets.

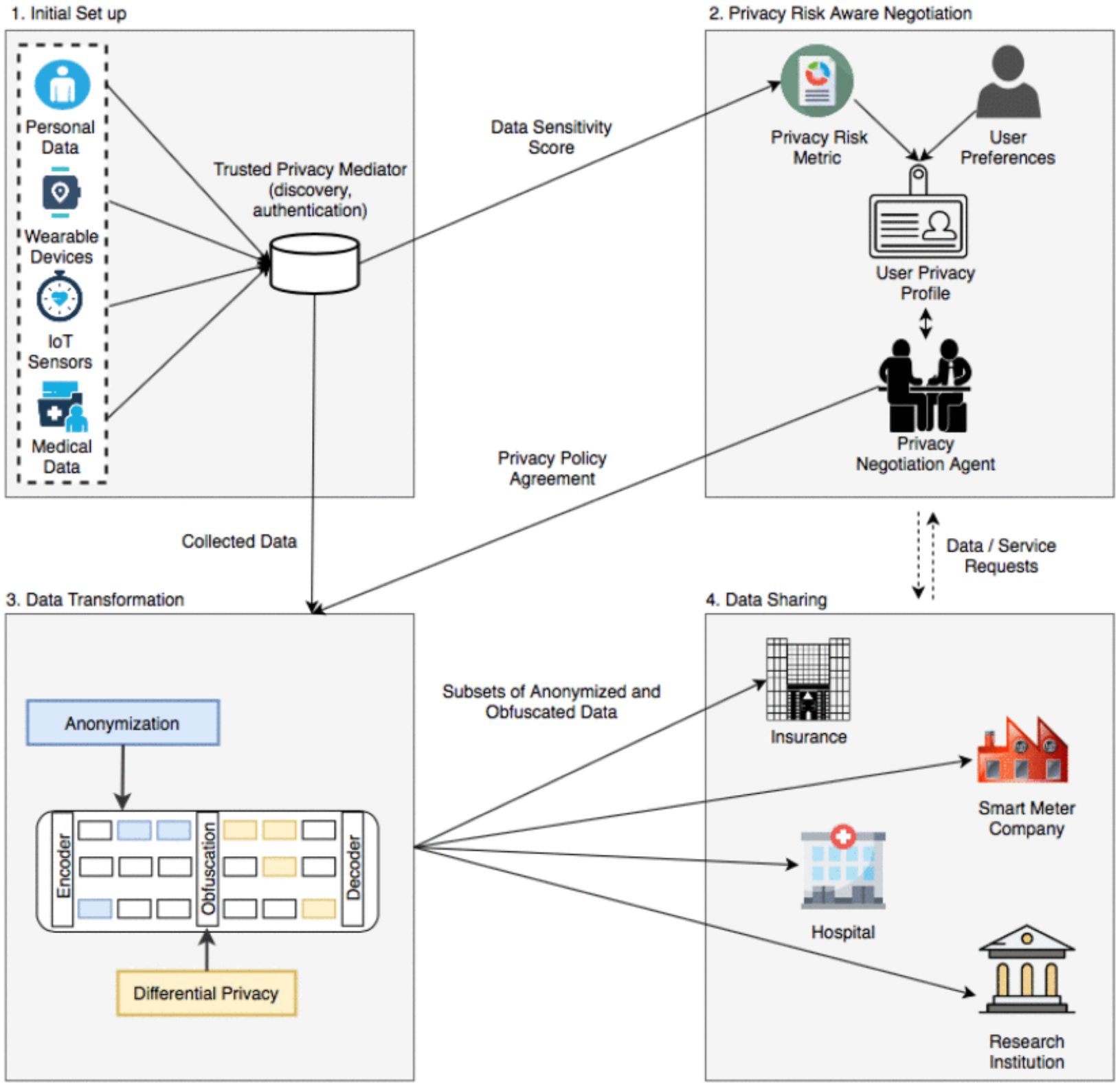

*Figure 13: **Example Network Analysis Architecture**. Detailed systems analysis can use existing proposed architectures as a starting point. Figure from [71].*

One possible path forward for providing objectives and resources for the above recommendations would be the establishment of a committee on the scientific survey of cyber defense to consider how far recent advances in scientific and technical knowledge can be used to strengthen the present methods of cyber defense. In addition, the observatories and outpost community, along the lines of other scientific communities, could establish a committee to develop a comprehensive research strategy and vision for the next decade of transformative science at the frontiers of this domain.

## Acknowledgements

The authors wish to acknowledge the following individuals for their contributions and support: Bob Bond, Alan Edelman, Jeff Gottschalk, Chris Hill, Charles Leiserson, Mimi McClure, Damian Menscher, Steve Rejto, Daniela Rus, Allan Vanterpool, Marc Zissman, and the MIT SuperCloud team: Bill Arcand, Bill Bergeron, David Bestor, Chansup Byun, Michael Houle, Matthew Hubbell, Anna Klein, Lauren Milechin, Julie Mullen, Antonio Rosa, Albert Reuther, Charles Yee.

## Appendix A: Better Data for a Better Internet - A Call to Facilitate and Encourage Data Sharing

The Internet is now a part of the critical infrastructure of society and is essential for industries of the future. Its users depend on the Internet for news and information, for communication, for access to services, as well as entertainment and education.

At the same time in most parts of the world, the Internet infrastructure is the product of the private sector. Its shape has largely been determined by the decisions of the private sector, and the economic considerations that drive that sector. These considerations shape the interconnected character of the Internet, its degree of resilience, key aspects of its security, concerns about privacy, and its overall future trajectory. As the Internet becomes critical to society, society needs a rigorous understanding of the Internet ecosystem, a need made more urgent by the rising influence of adversarial actors.

For areas of critical concern to society, such as health care, transportation safety, or food safety, the government plays a role that complements the role of the private sector—it monitors the state of those systems, and acts as necessary to ensure that they are meeting the needs of society. The first step in this process is gathering data to understand how the system is actually working. Today, operators, policy makers and citizens have no consensus view of the Internet to drive decision-making, understand the implications of current or new policies, or to know if the Internet is being operated in the best interests of society.

Governments could gather data directly, but the trans-national character of the Internet raises challenges for government coordination. An accepted approach to data gathering and analysis is to make sure that data is made available to neutral third-parties such as academic researchers, who can independently pursue their efforts, draw their own conclusions, subject these to comparison and peer review, and present their results as advice to governments.

Most of the data that can provide insights into the state of the Internet is in the hands of the private sector actors that implement and operate the Internet. Some can be gathered by independent measurement, probing the Internet from the edge, but much of it is best – or only – obtainable by the private sector actors themselves. Many questions about the security, stability, and resilience of critical infrastructure will require scientific cooperation between the private sector and academia, with the encouragement and support of governments. Topics of interest include: significant network outages; hijacking of routing and naming layers of the infrastructure; botnet source, scope, and spread; and persistent or recurring congestion and performance impairments.

The purpose of this open letter is to urge governments to put their support behind a program of cooperation between the private sector and independent researchers and experts to share data and support analysis, with the ultimate goal of giving governments and society a view as to how the Internet is serving its role as critical infrastructure.

A call to share data does not mean that the data has to be made public. There are well-understood practices, used both in this sector and in other sectors, to allow access to data by qualified independent scholars in a responsible manner. We outline some of those practices and principles in the appendix to this letter. What is necessary is that independent researchers can have access to data in a way that allows them to replicate or undertake original research, building on previous work. Currently, data-driven research concerning the Internet is often a one-time effort, perhaps using proprietary data that cannot be shared. The importance of the Internet requires that the research community move beyond this mode of operation to a more sustained, scientific engagement.

Sharing of data by the private sector is not without risk, which triggers understandable hesitation. If data contains personally identifiable information (PII), procedures must be put in place to protect privacy, and governments must affirm that those procedures are consistent with their laws and regulations. In the case of health care in the U.S., HIPAA (the Health Insurance Portability and Accountability Act) has specific provisions that govern research practices that use medical data. As well, release of certain data could conceivably lead to adverse commentary on some stakeholder, or policies adverse to the stakeholder's interests. We recognize all these concerns, but it has become clear that governments should encourage and participate in a solution to the tremendous counter-incentives to share data to support Internet science. In computer security the Menlo Report has proposed approaches to navigate these challenges in the case of specific incident and threat data.

We also see compensating benefits to the private sector in a program of increased data sharing. Each actor in the Internet ecosystem may have an accurate view of their part of the system, but may not have a similar understanding about the state of their competitors, or the larger ecosystem. Allowing neutral third parties to obtain data from multiple actors can give the private sector, as well as governments and society, a global view of the state of the Internet. Additionally, sharing of data in the way described here, with controls on further disclosure, would not preclude the opportunity for the private sector to make other use of that data, including for commercial purposes. But an important message from governments should be that responsible sharing of data for documented scientific research will not generate corporate liability.

There is another benefit to the policy we call for here: the academic training of professionals to work with large data sets focused on communications and networking. While synthetic network data can be used for classroom exercises, serious research of the sort that leads to professional development requires real data, with the genuine potential for new discovery. We are not calling for a single, gigantic, coordinated, or permanent data collection, but multiple data sets targeted at scientific understanding of existing infrastructure vulnerabilities and impacts of their exploitations.

We are making a request to governments that they in turn send a strong signal to the private sector that builds and operates the Internet: data sharing is a necessary aspect of sustaining critical infrastructure, the Internet has now reached this level of maturation, and (as is true in other aspects of society) responsible data sharing needs to be part of normal practice. Developing this model now is a worthwhile activity before some future Internet catastrophe forces an ad-hoc approach to Internet data sharing that would be less beneficial to operators, policymakers, and citizens.

## Appendix B: Communicating Data Release Requests

When communicating a data release request with an information security officer (ISO), the following topics should be kept in mind and touched upon in the initial communication. While it is good to have additional information available if follow-ups are requested, the initial communication should be kept fairly short and minimize the use of domain specific terminology.

What is the data you're seeking to share?

- Describe the data to be shared, focusing on its risk to the organization if it were to be accidently released to the public or otherwise misused.

Example: The data was collected on <<date range>> at <<location(s)>> in accordance with our mission. The risk has been assessed and addressed by an appropriate combination of excision, anonymization, and/or agreements. The release to appropriate legitimate researchers will further our mission and is endorsed by leadership.

Explanation: Sentence 1 establishes the identity, finite scope, and proper collection of the data. Sentence 2 establishes that risk was assessed and that mitigations were taken. Sentence 3 establishes the finite scope of the recipients, an appropriate reason for release, and mission approval.

Where / to whom is the data going?

- Please describe the intended recipients of the data, the systems they will use to receive / process the data.

Example: The data will be shared with researchers at <<institution>>. The data will be processed on <<institution>> owned systems meeting their institution security policies, which include password controlled access, regular application of system updates, and encryption of mobile devices such as laptops. Authorized access to the data will be limited to personnel working as part of this effort.

Explanation: Sentence 1 establishes the legal entity trusted with the data and with whom any agreements are ultimately made on behalf of. Sentence 2 establishes that basic technical safeguards are in place, without getting too specific, and that personally-owned computers will not be used as the institution has no legal control over them. Sentence 3 establishes that the data will not be used for other purposes than the agreed-upon research project.

What controls are there on further release (policy/legal & technical)?

- Is a non-disclosure or data usage agreement in place?
- Is the data anonymized? If so, is there an agreement in place to prohibit de-anonymization attempts?
- What technical controls are in place on the systems that will receive / process the data to prevent misuse?
- Is there an agreement in place on publication of results from this effort?
- Is there an agreement in place for the retention or deletion of the original data, intermediate products, and/or the results at the end of the effort?

Example: An acceptable use guidelines that prohibit attempting to de-anonymize the data and will be provided to all personnel working on the data. Publication guidelines have been agreed to that allow for high-level statistical findings to be published, but prohibit including any individual data records. A set of notional records has been provided that can be published as an example of the data format, but is not part of the actual data set. The research agreement requires all data to be deleted at the end of the engagement except those items retained for publication.

Explanation: Sentence 1 establishes that there is an agreement in place prohibiting de-anonymizing the data and clearly defining it as "misuse" of the data. Sentence 2 and 3 establish that it is known to all parties what may and may not be published. Sentence 4 establishes that data retention beyond the term of the agreement has been addressed and cleanup is planned as part of project closeout.